\documentclass[journal]{IEEEtran}
\usepackage{graphicx}

\begin{document}

\title{Superconducting Neuromorphic Computing Using Quantum Phase-Slip Junctions}

\author{Ran~Cheng,
        Uday~S.~Goteti,
        and~Michael~C.~Hamilton,~\IEEEmembership{Senior~Member,~IEEE}
\thanks{R. Cheng, U. S. Goteti, and M. C. Hamilton are with the Department
of Electrical and Computer Engineering, Auburn University, Auburn,
AL, 36849 (e-mail: mchamilton@auburn.edu).}}

\markboth{ASC2018-2EPo1A-11}%
{Shell \MakeLowercase{\textit{et al.}}: Superconducting neuromorphic computing using quantum phase-slip junctions}

\maketitle

\begin{abstract}
 Superconducting circuits based on quantum phase-slip junctions (QPSJs) can conduct quantized charge pulses, which naturally resemble action potentials generated by biological neurons. A corresponding synaptic circuit, which works as a weighted connection between two neurons, can also be realized by circuits comprised of QPSJs and magnetic Josephson junctions (MJJs) as a means of charge modulation for quantized charge propagation. In this paper, we present basic neuromorphic components such as neuron and synaptic circuits based on superconducting QPSJs and MJJs. Using a SPICE model developed for QPSJs, neuron and synaptic circuits have been simulated in WRSPICE to demonstrate possible operation. We provide estimates for QPSJ energy dissipation and operation speed based on calculations using simple models. The challenges for implementation of this technology are also briefly discussed.

\end{abstract}
\begin{IEEEkeywords}
quantum phase-slip junction, superconducting neuron, synapse, non-von Neumann.
\end{IEEEkeywords}

\IEEEpeerreviewmaketitle

\section{Introduction}
\IEEEPARstart{N}{euromorphic} computing is a brain-inspired non-von Neumann architecture that has been realized in CMOS technology \cite{mead1990neuromorphic,linares2011spike,seo201145nm,rajendran2016neuromorphic,davies2018loihi}. Mapping conventional neural networks running on a von Neumann machine to a neuromorphic platform designed specifically for neural networks can significantly reduce power consumption and improve processing efficiency \cite{braga2010performance}. Recently, new materials and devices, including transistors \cite{hu2017mos2}, organic electronics \cite{van2017non,van2018organic}, and memristive devices \cite{rajendran2016neuromorphic,jeong2018nonvolatile}, have been developed for realization of neurons and synapses in neuromorphic chips. However, due to the CMOS technology bottleneck, the major challenge of neuromorphic systems is the massive power dissipation, which is several orders of magnitude behind a human brain \cite{painkras2013spinnaker}. This situation becomes even worse when attempting to scale up to match the capabilities of a human brain.

On the other hand, superconducting electronics incorporating Josephson junctions (JJs) have been shown to be potential candidates for neuromorphic computing due to their ultra-low power dissipation and high-speed operation \cite{schneider2017energy}. The nearly loss-less property of massive interconnections in superconducting circuits makes them promising for lower interconnect loss compared to CMOS circuits with lossy interconnects. In a JJ-based neuromorphic system, single flux quantum (SFQ) pulses are recognized as action potentials and magnetic Josephson junctions (MJJs) \cite{bell2004controllable,ryazanov2012magnetic}, with tunable critical currents, can be designed to behave like synapses of a biological neural network \cite{russek2016stochastic}. Furthermore, quantum phase-slip events in one-dimensional superconducting nanowires have been observed as resistive tails in I-V characteristics below a critical temperature and above a critical voltage \cite{giordano1988evidence}. This phase-slip phenomenon has been identified as an exact dual to Josephson tunneling \cite{mooij2006superconducting}. The Josephson effect can be explained as charge tunneling across the dielectric barrier in a superconductor-dielectric-superconductor structure, while quantum phase-slip can be viewed as flux-quantum tunneling across the superconducting nanowire in a dielectric-superconductor-dielectric structure. We previously showed that, with appropriate configuration of a quantum phase-slip junction (QPSJ) circuit, a quantized charge current pulse can be generated and propagated \cite{goteti2018charge,goteti2018Comp}. This type of circuit can also emulate brain behaviors, possibly with even lower power dissipation, since the static power dissipation of a QPSJ is almost zero. The quantum phase-slip phenomenon has been experimentally demonstrated by multiple research groups \cite{mooij2006superconducting,constantino2018emergence}, but applications remain at a preliminary stage. Our QPSJ SPICE model \cite{goteti2015spice} is thus based on the theoretical model proposed by Mooij et al.\cite{mooij2006superconducting} and is well to simulate the circuits in this paper.    

We present components of a superconducting neuromorphic system comprised of basic neuron and synaptic circuits based on QPSJs and MJJs. In Section II, we introduce our QPSJ-based neuron circuit using SPICE simulation \cite{goteti2015spice}. The neuron circuit, which integrates quantized current input pulses onto a capacitor and fires a current pulse when the capacitor voltage is above a threshold, behaves like a digital integrate-and-fire neuron. Charge modulation in our synaptic circuit is achieved through critical current modulation of an MJJ, similar to \cite{schneider2018ultralow}. The circuit functions are demonstrated through simulations in WRSPICE. A simple 2-layer network comprised of neuron and synaptic circuits has been simulated to illustrate the basic neuromorphic function of our system. As an estimation, we provide the power dissipation estimation of each switching operation of a QPSJ, compared to theoretical calculations for a JJ. In Section III, we discuss some of the limitations of superconducting neuromorphic systems, as well as experimental challenges of QPSJ-based circuits.

\section{Design and simulation}
\subsection{Neuron Circuit}
A voltage-biased QPSJ is a dual equivalent of a current-biased JJ \cite{mooij2006superconducting}, with respect to the duality of phase and charge. A single QPSJ is a nonlinear device that can be modelled by an intrinsic junction in series with a normal-state resistor and an inductor. The operation of a single QPSJ can be treated as an RLC oscillator with a damping parameter given by:
\begin{equation}
\beta_L=\frac{2\pi V_cL}{2eR^2}
\end{equation}
where $\beta_L$ is the damping parameter, $V_c$ is the critical voltage of the QPSJ, $L$ is the inductance and $R$ is the normal resistance. The QPSJ is overdamped if $\beta_L \ll 1 $ and underdamped if $\beta_L \gg 1$. Similar to an overdamped JJ circuit that can produce a single flux quantum (SFQ) pulse, an overdamped QPSJ circuit can generate a quantized current pulse that contains a charge of $2e$, as we demonstrated through WRSPICE simulation in our previous work \cite{hamilton2018superconducting,goteti2018charge,cheng2018spiking}.

In the neuron circuit shown in Fig. 1, all the QPSJs are identical and biased by a voltage source $V_b$ \cite{cheng2018spiking}. The bias voltage usually provides $\sim$ 70\% of the critical voltage for each of these QPSJs. Therefore, all the QPSJs exhibit Coulomb blockade \cite{webster2013nbsi}, resulting in near zero current through each junction. The resistor $R_b$ is used to balance the bias voltages of $Q_0$ and multiple parallel QPSJs, i.e. $Q_1$ to $Q_N$. Upon arrival of a short voltage pulse from $V_{in}$, $Q_0$ is switched and generates a current pulse that contains a charge of $2e$. The capacitor $C$ is selected to store a charge $Q = 2Ne$. The voltage on the capacitor increases by an amount of $2e/C$ every time a current pulse from $Q_0$ is injected to $C$, until the voltage across the parallel QPSJs is above their critical voltage. At that point, a $2e$ current pulse is generated at each of the QPSJs and sums up at node 2. This circuit works similar to a digital integrate-and-fire neuron, which counts the number of input spikes and fires a spike after reaching a pre-defined threshold. Here the number of parallel QPSJs $N$ defines the firing threshold. 

\begin{figure}[t]
\centering
\includegraphics[scale=0.32]{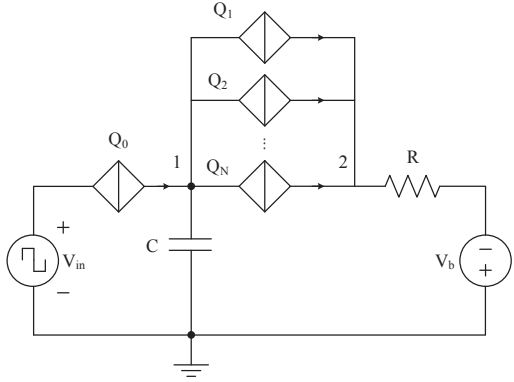}
\caption{An example QPSJ-based neuron circuit.}
\end{figure} 

Fig. 2 shows the simulation results of a QPSJ neuron circuit with threshold $N$ equal to 10. Input rectangular voltage pulses have a period of 100 ps and a width of 3ps. The voltage at capacitor $C$ keeps increasing as a result of quantized charge accumulation. Once the voltage applied on the parallel QPSJs reaches the critical voltage, the capacitor immediately discharges $20e$ charge. Each $2e$ charge is transported through a parallel QPSJ, resulting in a spike-like current pulse. We can also observe that the magnitude of current pulse through $Q_0$ decreases periodically, because the voltage across $Q_0$ decreases when the voltage at the capacitor keeps increasing. The input voltage pulse width and magnitude must be appropriately selected so that a single quantized current pulse can be generated during the voltage integration process. Otherwise, multiple pulses might be generated after one input voltage pulse.

\begin{figure}[t]
\centering
\includegraphics[scale=0.32]{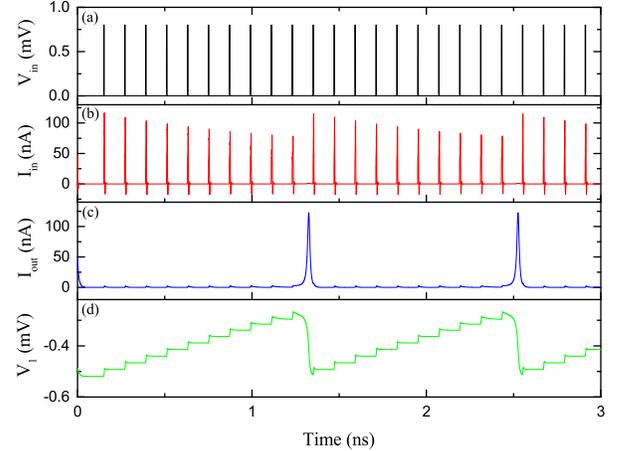}
\caption{Simulation results of neuron circuit shown in Fig. 1. Input are voltage pulses with 0.8 mV magnitude, 3 ps width, and 120 ps period. Critical voltage of each QPSJ is 0.7 mV and $V_b$ is 1 mV. Normal resistance of $Q_0$ is 10 k$\Omega$, normal resistance of $Q_1$ to $Q_N$ is 15 k$\Omega$ and resistance of $R$ is 9 k$\Omega$. (a) Input voltage. (b) Input current. (c) Output current. (d) Capacitor voltage at node 1.}
\end{figure} 

\subsection{Synaptic Circuits}
In a brain, a synapse connects two neurons and determines the signal strength transmitted from a pre-synaptic neuron to a post-synaptic neuron. Similarly, a synaptic circuit should be able to adjust the connection strength between two neuron circuits. In CMOS neuromorphic systems, non-volatile memory cells are usually used to implement synaptic circuits \cite{jeong2018nonvolatile}. However, a lack of non-volatile superconducting devices/circuits made superconducting neuromorphic implementation more challenging, until the recent realization of magnetic Josephson junctions (MJJs) for this purpose \cite{bell2004controllable,ryazanov2012magnetic}. An MJJ has a tunable critical current that can control the switching threshold \cite{schneider2017energy} to function as a binary synapse or control the circulating current in a superconducting loop to function as an analog synapse \cite{schneider2018ultralow}. Although a corresponding tunable critical voltage of a QPSJ has not yet been theoretically or experimentally demonstrated, we were able to combine MJJs and QPSJs to realize synaptic functions, as illustrated in the following paragraphs.

\begin{figure}[t]
\centering
\includegraphics[scale=0.32]{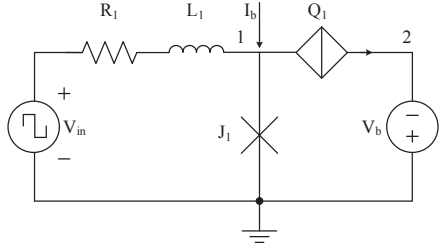}
\caption{A binary synaptic circuit based on a superconducting MJJ ($J_1$) and QPSJ ($Q_1$).}
\end{figure} 

\begin{figure}[t]
\centering
\includegraphics[scale=0.32]{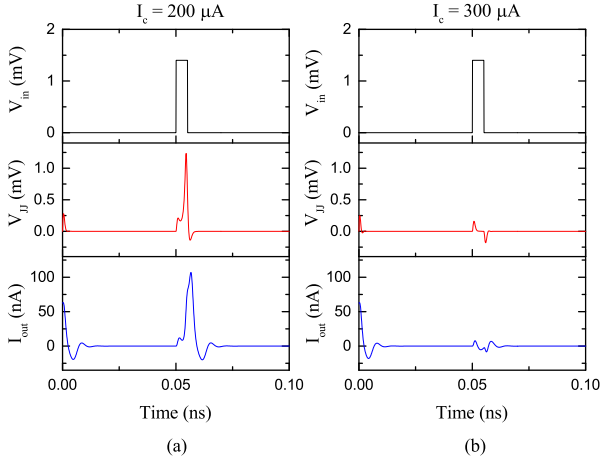}
\caption{Simulation results of input voltage, voltage at $J_1$, and output current in a binary synaptic circuit with different $J_1$ critical current $I_c$. Critical voltage of $Q_1$ is 0.7 mV, bias current $I_b$ is 140 $\mu A$, and magnitude of input pulse $V_{in}$ is 1.4 mV. (a) $I_c$ is 200 $\mu A$. (b) $I_c$ is 300 $\mu A$.}
\end{figure} 

A simple binary synaptic circuit using an MJJ and a QPSJ is shown in Fig. 3. Initially, MJJ $J_1$ is biased by a DC current $I_b$ while QPSJ $Q_1$ is biased by $V_b$. A short voltage pulse from input $V_{in}$, through $R_1$ and $L_1$, provides a current pulse for $J_1$. Inductor $L_1$, which can be omitted in this circuit operation, is used to trap an SFQ pulse from previous stage if there are multiple JJ stages. If $J_1$ is designed to have two distinct critical currents controlled by the magnetic order parameters \cite{russek2016stochastic}, the junction can be either switched or not upon the arrival of an input current pulse. For instance, in the case of low critical current of $J_1$, the input current pulse is large enough to switch $J_1$. An SFQ pulse is generated at node 1, which can, in turn, switch $Q_1$, resulting in a current pulse at the output. However, if the critical current of $J_1$ is high, the total current injected to $J_1$ is insufficient to switch it. Voltage at node 1 stays at zero and there will be no quantized current pulses at the output. The full operation of this binary synaptic circuit is illustrated in the simulation results shown in Fig. 4, where the critical current of $J_1$ is either 200 $\mu A$ or 300 $\mu A$, to represent a weight of "1" or "0". The operation of a network comprising QPSJ neuron circuits and synaptic circuits will be discussed in Section II. C. 

In the neuromorphic system described here, signals are generated and propagated in the form of quantized charge current pulses. The multi-state synaptic circuit shown in Fig. 5 can generate multiple current pulses that contain a charge of $2Ne$, depending on the state of MJJ $J_2$. Josephson junction $J_1$ is biased by $I_b$ and switched after an input current pulse from $V_{in}$. Inductor $L_2$ is chosen to store a single flux quantum, which in turn switches MJJ $J_2$ after a short delay. The resulting voltage pulse at node 2 switch $Q_1$ and generate current pulses at output. Since the magnitude of SFQ pulse at node 2 is inversely proportion to the critical current $I_c$ of $J_2$, the width of this SFQ pulse is proportion to $I_c$. As a result, more current pulses are generated when $I_c$ is low because the switching speed of $Q_1$ is fast (i.e. the normal resistance of $Q_1$ is relatively low), which allows $Q_1$ to be switched multiple times within a short time. The simulation results show that different numbers of current pulses are generated when $I_c$ varies from 10 $\mu A$ to 400 $\mu A$. Although the change is not linear, this can be used to represent a synaptic weight of 2, 1, 0.5, or 0, according to the number of current pulses generated within a set time interval.

\begin{figure}[t]
\centering
\includegraphics[scale=0.32]{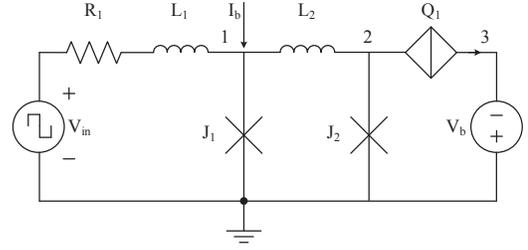}
\caption{A multi-state synaptic circuit based on a superconducting JJ ($J_1$), MJJ ($J_2$), and QPSJ ($Q_1$).}
\end{figure} 

\begin{figure}[t]
\centering
\includegraphics[scale=0.32]{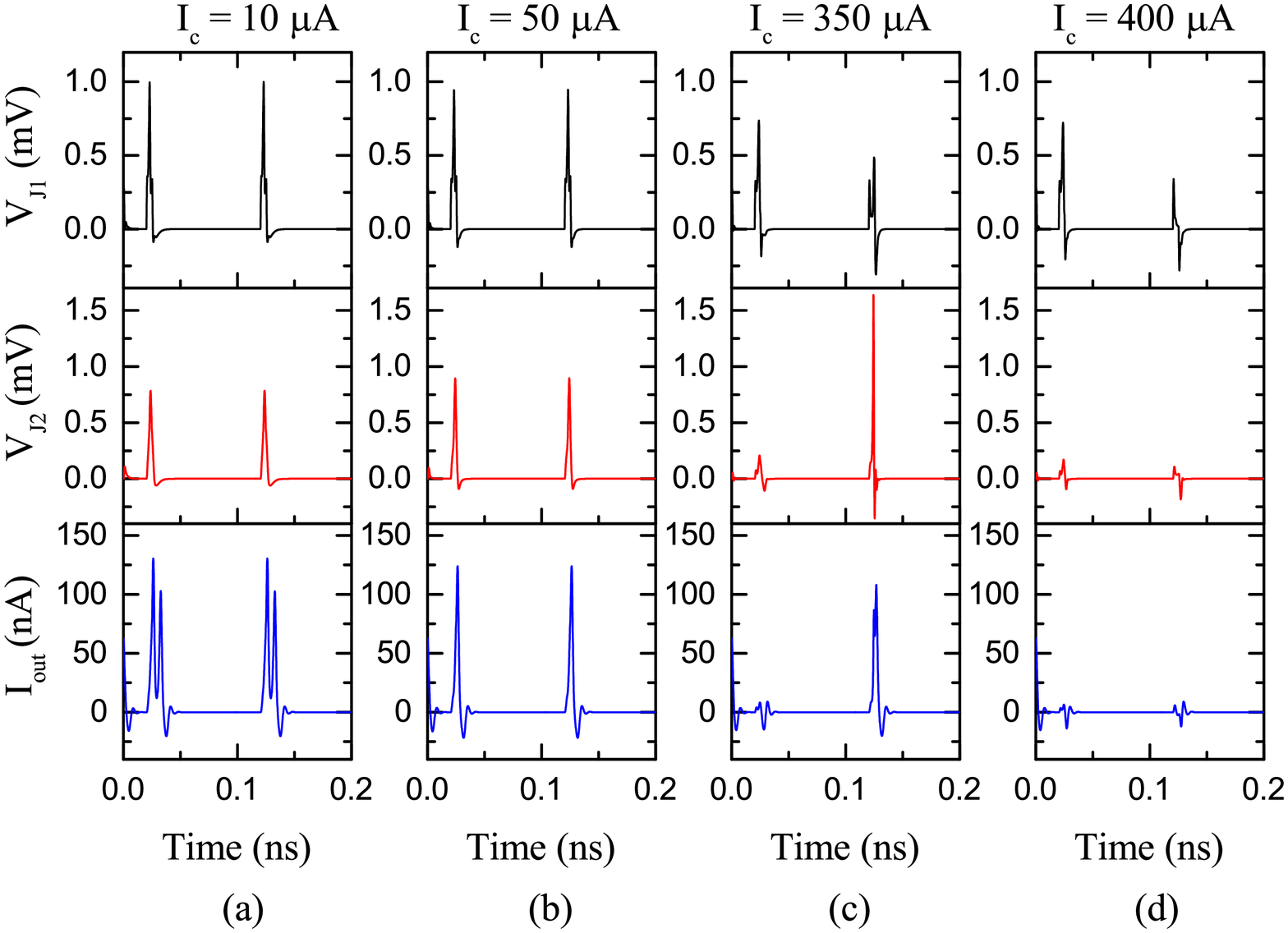}
\caption{Simulation results of voltage at $J_1$, voltage at $J_2$, and output current in a multi-state synaptic circuit with different $J_1$ critical current $I_c$. Critical voltage of $Q_1$ is 0.7 mV, critical current of $J_1$ is 200 $\mu A$, and bias current $I_b$ is 160 $\mu A$. (a) $I_c$ is 10 $\mu A$. (b) $I_c$ is 50 $\mu A$. (c) $I_c$ is 350 $\mu A$. (d) $I_c$ is 400 $\mu A$.}
\end{figure} 

\subsection{Neural Network Simulation}
\begin{figure}[t]
\centering
\includegraphics[scale=0.25]{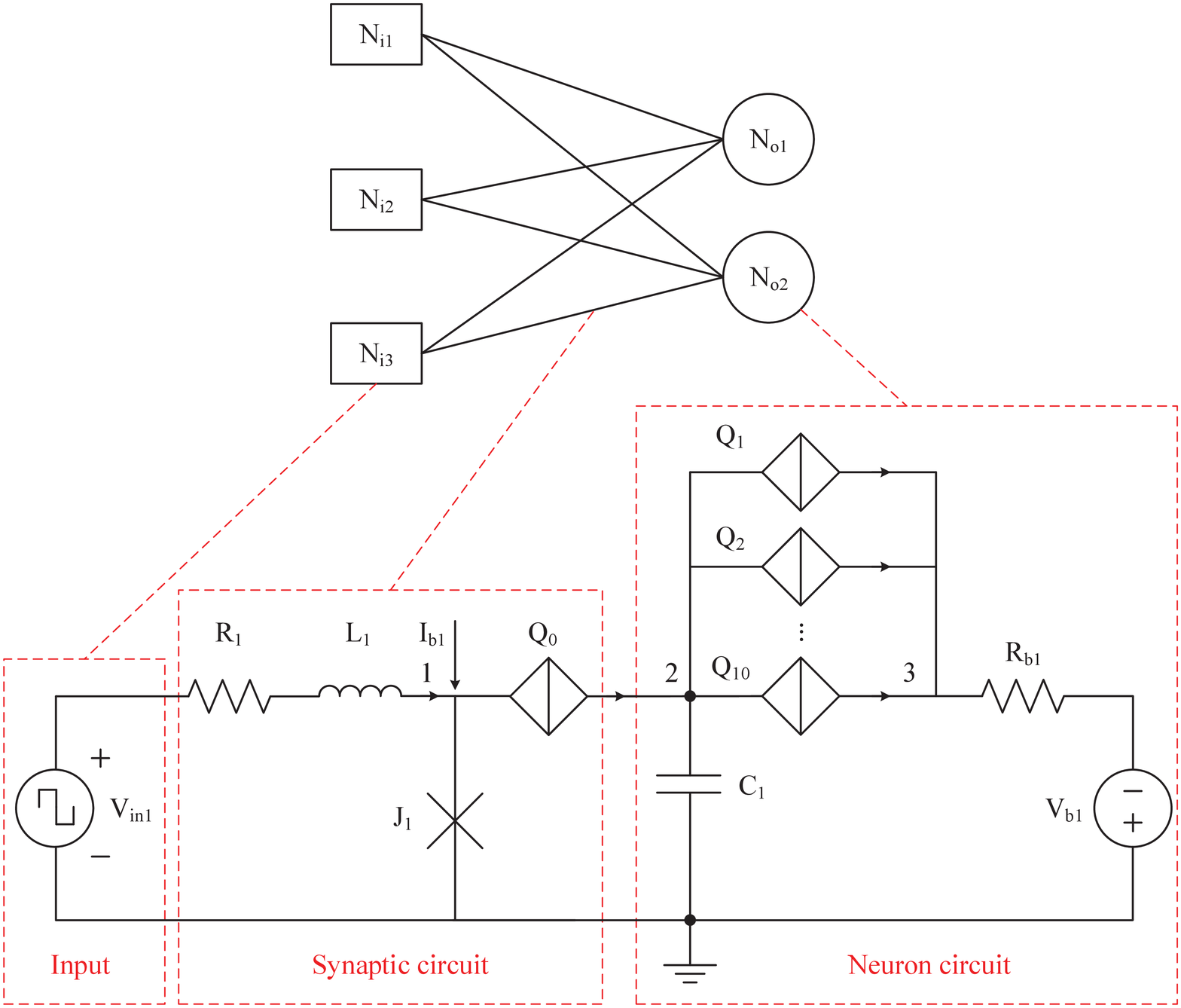}
\caption{Network architecture of a 3$\times$2 neural network based on superconducting synaptic and neuron circuits. $N_{ix}$ are input neurons and $N_{oy}$ are output neurons.}
\end{figure}

\begin{figure}[t]
\centering
\includegraphics[scale=0.32]{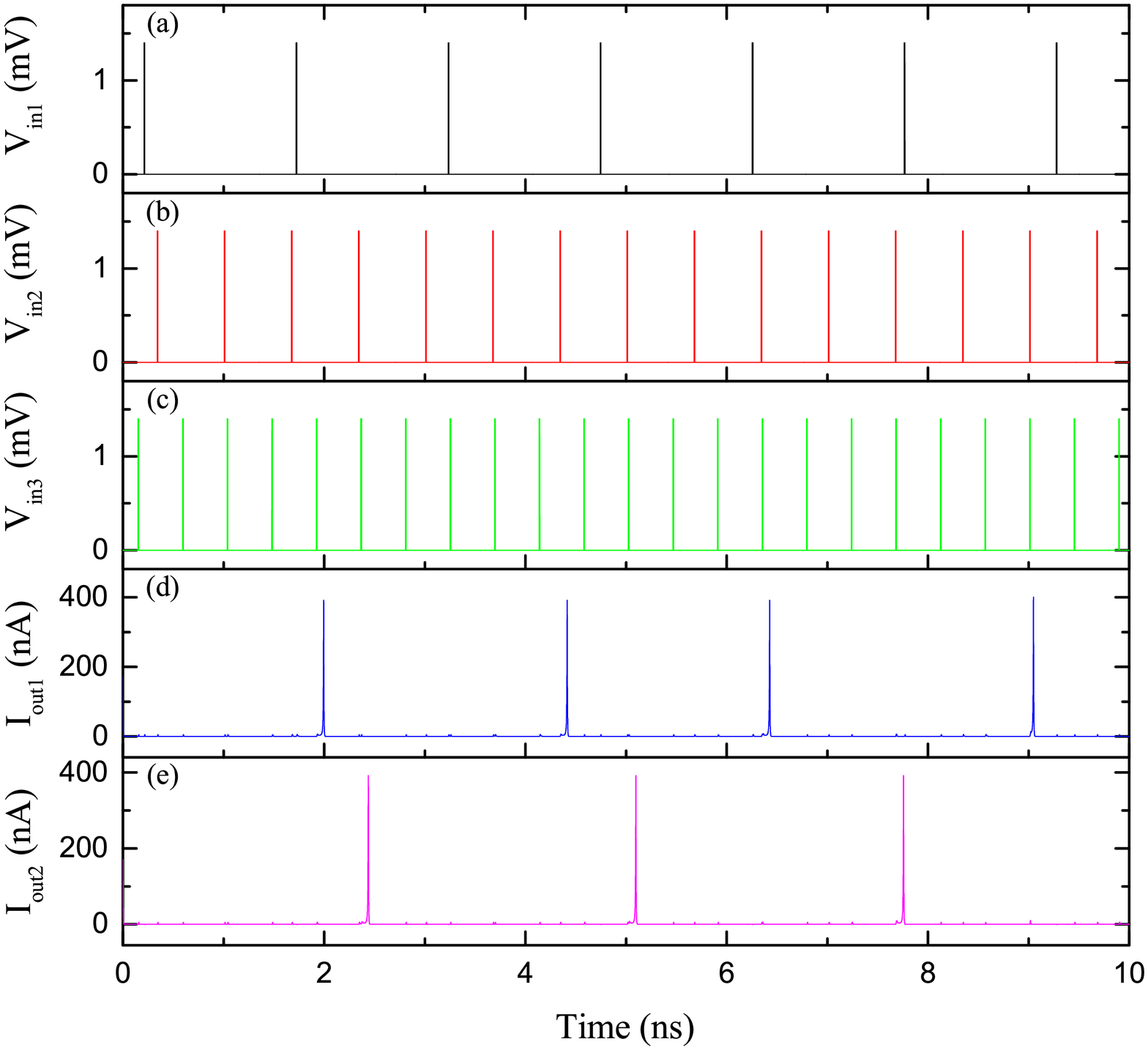}
\caption{Simulation results of a 3$\times$2 network illustrated in Fig. 7 with weight matrices [1 1 1] and [0 1 1]. (a) Input voltage 1. (b) Input voltage 2. (c) Input voltage 3. (d) Output current 1. (e) Output current 2.}
\end{figure}

\begin{figure}[t]
\centering
\includegraphics[scale=0.32]{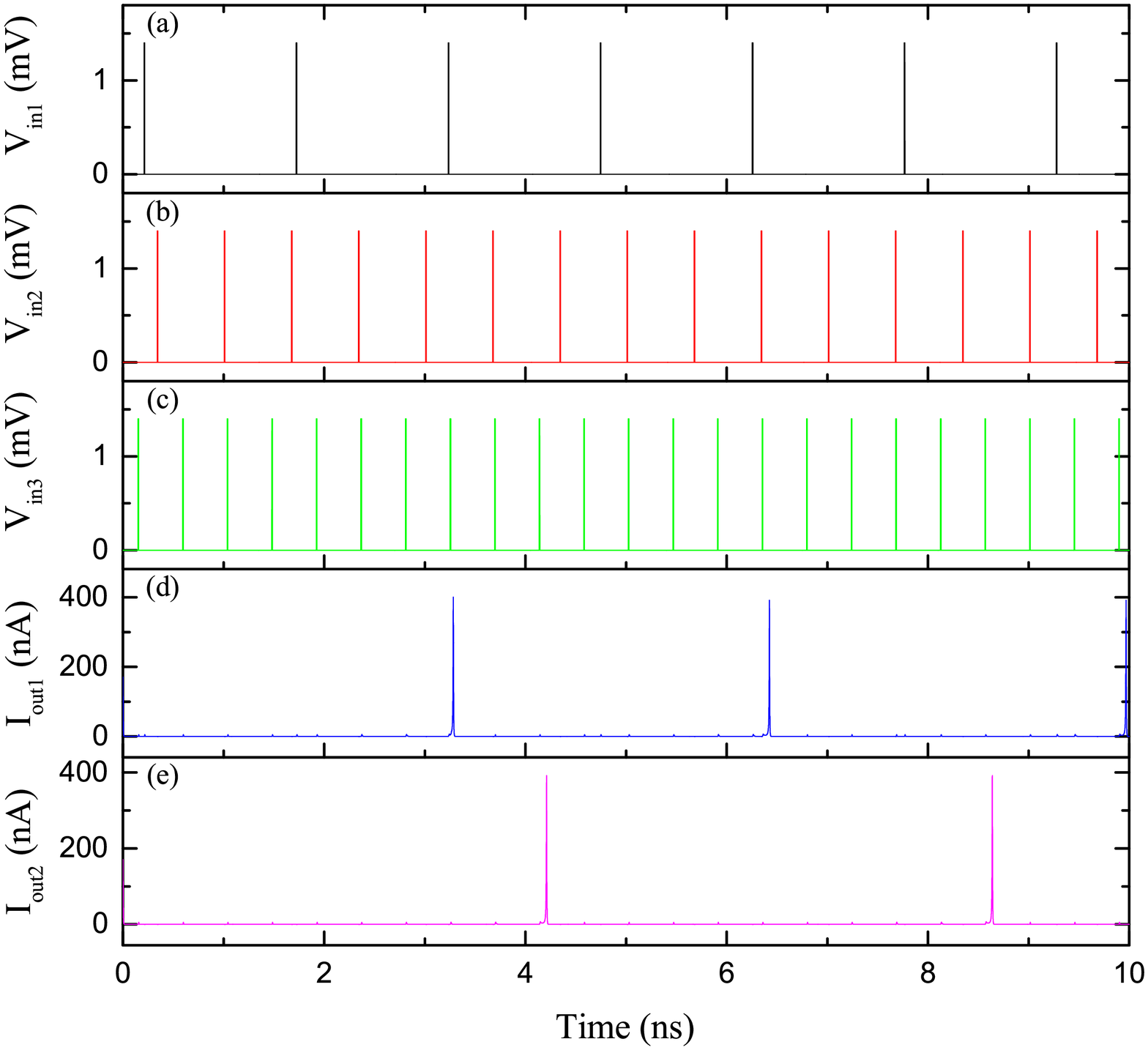}
\caption{Simulation results of a 3$\times$2 network illustrated in Fig. 7 with weight matrices [1 0 1] and [0 0 1]. (a) Input voltage 1. (b) Input voltage 2. (c) Input voltage 3. (d) Output current 1. (e) Output current 2.}
\end{figure}

To verify the functions of our neuron and synaptic circuits and demonstrate extension to more complex circuits, we combined neuron and synaptic circuits and simulated a neural network. A basic 3$\times$2 network architecture is shown in Fig. 7, for simplicity. We simulated the 3$\times$2 network consisting of three voltage sources and two neuron circuits connected by six binary synaptic circuits. Voltage sources provide input pulses with different frequencies, which represent three different external signals or signals from upstream neurons. The output neuron fires a pulse/spike as soon as it receives a set number (in this case 10) of quantized current pulses from the three synaptic circuits, as the threshold is set to 10. The simulation was conducted with different initial values of each synaptic weight (i.e. critical current of $J_1$), that reflected different firing frequencies observed for each output neuron. Weight matrices for neuron 1 and neuron 2 were [1 1 1] and [0 1 1], respectively, in Fig. 8, and were [1 0 1] and [0 0 1], respectively, in Fig. 9. The results show that pulses from a weight "1" synapse is counted while those from a weight "0" are not counted, and the neuron fires immediately after the number of total pulses received reaches the threshold (= 10 here). As a result, this circuit works similar to a digital neuromorphic circuit. If we replace the synaptic circuit in Fig. 7 with a multi-state synaptic circuit, the network will be more complex, but also, more flexible. 

\section{Discussion}
In this section, we briefly discuss power dissipation and design and experimental challenges in order to include limited comments on these topics, but by no means a complete discussion, which is outside the scope of this paper. The power dissipation of QPSJ-based circuits is expected to be extremely low, compared to CMOS circuits, and possibly even compared to JJ-based superconducting circuits. Estimates of the power dissipation and switching speed of a JJ and QPSJ can be calculated through equations given in \cite{kadin1999introduction,mooij2015superconductor}. As an estimation, the energy per switching event for a JJ circuit is approximately 0.33 aJ \cite{schneider2018ultralow}. For a practical QPSJ using InOx material, we calculated the energy per switching event for a QPSJ circuit as $\sim$ 3.2 zJ \cite{cheng2018spiking}. Therefore, the neuron circuit shown in Fig. 1 (threshold is 10) dissipates $\sim$ 35.2 zJ for a firing event. Furthermore, the switching speed of a QPSJ can be as low as several ps, depending on the particular material and device geometry.

Nevertheless, like many other superconducting neuromorphic systems \cite{schneider2018ultralow}, challenges remain for on-chip learning functions, which is important for a self-learning neuromorphic system. Although magnetic Josephson junctions are non-volatile devices, additional learning and control units are required to realize on-chip learning. Regardless of the writing efficiency of an MJJ, the present writing process requires an external magnetic field, which is not suitable for real-time learning circuit operations. To scale up this network, fan-in, fan-out, and biasing issues are also challenges for these types of devices, as evident in JJ circuits \cite{suzuki19894k}. As we mentioned before, QPSJ technology remains at a preliminary stage. Existing experimental issues for single-electron devices, such as charge fluctuations and charge offset \cite{likharev1999single,takahashi2002silicon}, are also expected to be potential challenges for QPSJs, with suitable experimental approaches still to be determined. Furthermore, as mentioned in Section II. B, an electric-field-tunable QPSJ has not yet been experimentally demonstrated. Predication of a device such as this would greatly enhance options for QPSJ-based neuron and synaptic circuits. 

\section{Conclusion}
Quantum phase-slip junctions are an emerging superconducting technology. We have presented superconducting neuromorphic circuits based on QPSJs. The intrinsic spiking behavior of QPSJ-based circuits makes it possible to emulate the action potentials of a biological neuron. Circuit functions have been demonstrated through simulations in WRSPICE. Furthermore, the estimated power dissipation of QPSJ-based circuits is promising, compared to other technologies. On-chip learning and large-scale integration of superconducting neuromorphic systems, along with experimental challenges, are presently obstacles for real applications of this type of circuits. We look forward to finding solutions to these challenges through our future research.

\ifCLASSOPTIONcaptionsoff
  \newpage
\fi

\bibliographystyle{IEEEtran}
\bibliography{bibli}

\end{document}